\documentclass[preprint]{aastex}
\shorttitle{Delay-Spectrum Foreground Removal}
\shortauthors{Parsons, et al.}

\usepackage{amsmath}
\usepackage{graphicx}
\usepackage{natbib}
\begin{document}
\title{A Per-Baseline, Delay-Spectrum Technique for Accessing the 21cm Cosmic Reionization Signature}

\author{Aaron R. Parsons\altaffilmark{1}, Jonathan C. Pober\altaffilmark{1}, 
James E. Aguirre\altaffilmark{2}, Christopher L. Carilli\altaffilmark{3}, Daniel C. Jacobs\altaffilmark{4}, and
David F. Moore\altaffilmark{2}}

\altaffiltext{1}{University of California, Berkeley}
\altaffiltext{2}{University of Pennsylvania}
\altaffiltext{3}{National Radio Astronomy Observatory, Socorro}
\altaffiltext{4}{Arizona State University}

\begin{abstract}
A critical challenge in measuring the
power spectrum of 21cm emission from cosmic reionization 
is compensating for the frequency dependence of an interferometer's sampling pattern, which
 can cause smooth-spectrum foregrounds to appear unsmooth and degrade the
separation between foregrounds and the target signal.  
In this paper, we present
an approach to foreground removal that explicitly accounts for this frequency dependence.
We apply the delay transformation introduced in
\citet{parsons_backer2009} to each baseline of an interferometer to concentrate
smooth-spectrum foregrounds within the bounds of the maximum geometric delays
physically realizable on that baseline. By focusing on
delay-modes that correspond to image-domain regions beyond the horizon,
we show that it is possible to avoid the bulk of smooth-spectrum foregrounds.
We map
the point-spread function of delay-modes to $k$-space, showing that delay-modes
that are uncorrupted by foregrounds also represent samples of the three-dimensional
power spectrum, and can be used to constrain cosmic reionization.
Because it uses only
spectral smoothness to differentiate foregrounds from the targeted 21cm signature, 
this per-baseline analysis approach relies on spectrally- and spatially-smooth
instrumental responses for foreground removal.  For sufficient levels of instrumental smoothness 
relative to the brightness of interfering foregrounds, this technique substantially
reduces the level of calibration previously thought necessary
to detect 21cm reionization.  As a result, this approach places fewer constraints on 
antenna configuration within an array, and in particular, facilitates
the adoption of configurations that are optimized for power-spectrum sensitivity.
Under these assumptions, we demonstrate the potential for the Precision Array for Probing
the Epoch of Reionization (PAPER) to detect 21cm reionization at an amplitude of
10 mK$^2$ near $k\sim0.2h\ {\rm Mpc}^{-1}$
with 132 dipoles in 7 months of observing.
\end{abstract}


% --------------------------------------------------------------------------
\section{Introduction}

Over the past decade, our theoretical understanding of how the Epoch of Reionization (EoR)
can be observed via the 21cm transition of neutral hydrogen
has advanced considerably as a result of the growing sophistication of numerical and
semi-analytic models of how ultraviolet and x-ray sources feed ionization fronts 
in the intergalactic medium \citep{furlanetto_et_al2006,wyithe_loeb2004,santos_et_al2010,mesinger2010,morales_wyithe2010,zahn_et_al2011}.
Three-dimensional tomographic imaging of 
temperature fluctuations as they grow and are erased by ionization will require a
collecting area comparable to a full Square Kilometer Array (SKA). However,
a statistical detection of the 21cm reionization signal through power spectrum analysis should be
 possible with $\sim 0.1\%$ of a square kilometer of collecting area. 
Many first-generation experiments aiming to measure the 21cm power spectrum are under
construction or already observing, including
the Giant Metre-wave Radio Telescope (GMRT;
\citealt{pen_et_al2009})\footnote{\url{http://gmrt.ncra.tifr.res.in/}},
the LOw Frequency ARray (LOFAR; \citealt{rottgering_et_al2006})\footnote{\url{http://www.lofar.org/}},
the Murchison Widefield Array (MWA; \citealt{lonsdale_et_al2009})\footnote{\url{http://www.mwatelescope.org/}}, and the Precision
Array for Probing the Epoch of Reionization (PAPER; \citealt{parsons_et_al2010})\footnote{\url{http://eor.berkeley.edu/}}.

An interferometer accesses the three-dimensional power spectrum of 21cm EoR emission by
measuring variation perpendicular to the line of sight
using samples provided by different baselines in the $uv$-plane,
and variation parallel to the line
of sight using the Fourier transform of frequency data \citep{morales2005}.
One of the major complications in using an interferometer to measure the
three-dimensional power spectrum of reionization is the frequency dependence of
the transverse wavemode sampled by a pair of antennas. This frequency dependence
causes smooth-spectrum foregrounds to appear unsmooth, as each frequency probes fluctuations
on different scales.
This effect degrades the
separation that can be achieved between foregrounds and higher spatial Fourier-domain $k$-modes of the
spherically averaged 21cm reionization power spectrum $\Delta_{21}^2(k)$, as foreground removal techniques generally rely
on the spectral smoothness of foreground emission to isolate the 21cm signal
\citep{morales_hewitt2004}. To overcome this difficulty in separating smooth-spectrum
foregrounds from the unsmooth 21-cm reionization signal, it has been
suggested that interferometric arrays should produce overlapping $uv$-coverage at
multiple frequencies \citep{bowman_et_al2009}, 
so that the same transverse Fourier mode is measured across the observing bandwidth. 
Generating such $uv$-coverage requires
large numbers of antennas, creating technical and logistical challenges that
are still in the process of being addressed for current interferometric arrays.
Moreover, combining samples of the same $uv$-pixel from different baselines at multiple
frequencies in a way that does not introduce spectral structure at the level of the expected
21cm EoR signal can pose daunting calibration challenges.

In this paper, we present
a novel approach to managing smooth-spectrum foreground contamination that explicitly accounts for the frequency dependence of the wavemodes
sampled by a baseline in an interferometer. We apply the delay transform described in
\citet{parsons_backer2009} (hereafter, PB09) separately to each baseline's visibility spectrum to concentrate
smooth-spectrum foregrounds within the bounds of the maximum delays
geometrically realizable on these baselines.  We show that by focusing on
modes that correspond to regions in image-domain that are beyond the horizon,
it is possible to avoid smooth-spectrum foregrounds altogether.  
This technique has several important advantages over traditional approaches.  Since each baseline
of the interferometer can be used to provide independent measurements of the power spectrum that are
free from foreground contamination, there is no need for longer baselines to aid in characterizing 
and removing foregrounds.  Instead, arrays utilizing the delay-spectrum approach can be deployed
using ``maximum-redundancy" array configurations (\citealt{parsons_et_al2012}, hereafter referred to as P12), reducing the demands on antenna number, correlator size, and image-domain sidelobe minimization 
for an array of comparable
sensitivity using more traditional antenna configurations.  Furthermore, the requirements on
calibration accuracy are reduced, with more emphasis put on limiting how rapidly responses evolve versus
frequency than on the level of absolute calibration of the array and its antenna elements.
With the delay-spectrum technique, we find that a substantial
window of opportunity is opened for measuring $\Delta_{21}^2(k)$
at spatial wavemodes of $k\gtrsim0.2h\ {\rm Mpc}^{-1}$.

In \S\ref{sec:dly_transform}, we present a description of the delay transform,
showing both how smooth-spectrum foregrounds are isolated in delay-space,
and that for shorter baselines, the Fourier modes obtained through the delay transform are 
highly peaked toward particular $k$-modes, making them effective measurements of the 21cm EoR power spectrum.  In \S\ref{sec:implementation}, we describe several important
details of how the delay transform is applied to spectra measured on each baseline of
an interferometer to effectively isolate foregrounds with minimal sidelobe contamination in the region
of interest for EoR.  In \S\ref{sec:simulation} we apply this technique to 
simulated visibilities
that incorporate realistic foreground properties, instrumental responses, and data flagging.
In \S\ref{sec:results} we present the results of our study of the delay transform, and 
in \S\ref{sec:discussion} we discuss the
implications of the delay transform for instrument design relative to other foreground mitigation techniques.
We conclude in \S\ref{sec:conclusion} with
prospects for applying this technique to actual observations.

\section{The Delay Transform}
\label{sec:dly_transform}

For interferometric measurements targeting the 21cm power spectrum of neutral hydrogen,
there are two ways to interpret spectral frequency.
For Galactic synchrotron and extragalactic point sources (i.e. ``foreground" emission), the frequency axis 
reveals the intrinsic broad-band spectrum of emission.  For 21cm emission,
however, frequency corresponds to the cosmological redshift of a spectral line, and hence,
a line-of-sight distance.  Actual measurements,
of course, contain both types of emission.  We
explore the relationship between these two interpretations of spectral frequency, paying particular
attention to result of applying the Fourier transform to the frequency spectrum that a single
baseline measures along the frequency direction.
The ultimate conclusion of this work
is that this Fourier transformation --- the ``delay transform" presented in PB09 --- 
localizes foreground emission in ``delay" space,
providing access to 21cm emission in regions uncontaminated by foregrounds.  Furthermore, we show
that for short baselines ($\sim 30$ meters) and small bandwidths ($\sim 8$ MHz), 
measurements of 21cm emission in delay space can be simply
interpreted as modes of the 21cm power spectrum.

\begin{figure}\centering
    \includegraphics[width=.95\columnwidth]{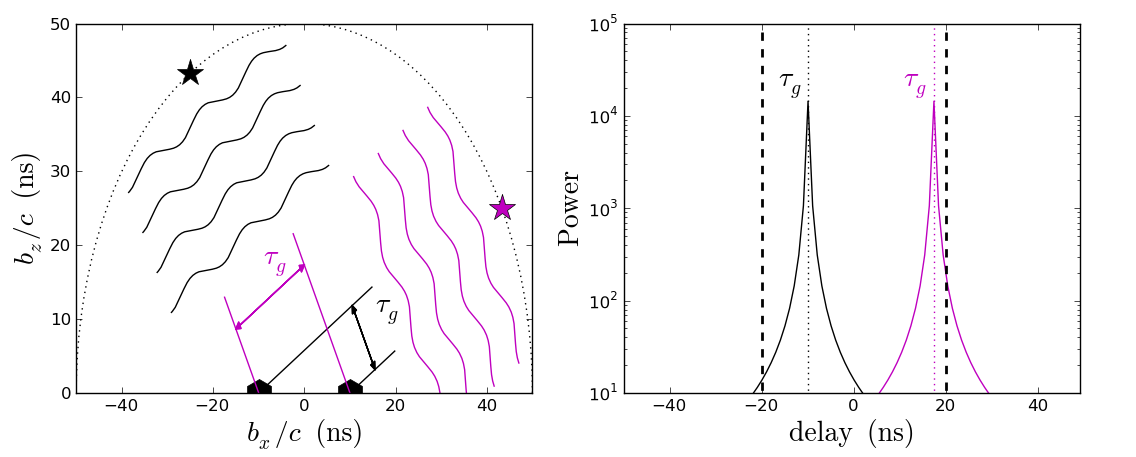}
\caption{
The geometric interpretation of the 
delay spectrum measured by an interferometer.  
The left plot shows how two sources with identical spectra can have differing geometric
delays $(\tau_g)$ owing to 
their positions relative to
the two antennas being correlated.  The right plot shows how a strictly geometric interpretation
of a delay spectrum is violated by the fact that the Fourier transform of the spectrum of each source
also enters the delay spectrum centered at the appropriate $\tau_g$. This is a manifestation of the convolution
expressed in equation \ref{eq:dly_t}.
Thick dashed lines denote the maximum possible geometric delays, as imposed by the horizon.  
While
the geometric delay associated with a source cannot exceed the horizon limit, sidelobes associated with the 
delay transform of the source spectrum do (right plot, magenta).
}\label{fig:spec_dly_shift}
\end{figure}

The ``delay transform" refers to the Fourier transform
of the visibility spectrum measured by a single baseline along the frequency axis;
a delay transform takes a single time sample of a frequency-spectrum of visibilities
from one baseline and Fourier transforms it to obtain 
a ``delay spectrum."
As shown in PB09, this transform relates the spectral frequency domain to the delay  (or ``lag'')
domain.  In delay domain, a flat-spectrum signal that arrives at one antenna
time-shifted by a delay $\tau$ relative to another antenna appears as a
Dirac delta function, $\delta_{\rm D}(\tau)$, in the delay spectrum of the
baseline pairing those antennas.  Because emission from different regions of
the sky arrive at antennas with different relative delays, bins in delay domain
can be used to geometrically select arcs on the sky.  For
interferometric arrays with wide fractional bandwidths, such as many
low-frequency 21cm reionization arrays, the delay transform can be a very
useful tool for isolating sources.  It effectively uses the
frequency dependence of a baseline's sampling of the $uv$-plane to make
one-dimensional images of the sky, as illustrated in Figure
\ref{fig:spec_dly_shift}.

However, in the context of measuring cosmic reionization using highly
redshifted 21cm emission from neutral hydrogen, the mapping between spectral
frequency and the line-of-sight direction makes the
Fourier transform of visibilities along the frequency direction an integral step in
measuring the three-dimensional power spectrum of reionization
$P_{21}(k)$ \citep{morales_hewitt2004,parsons_et_al2012}.  In this context,
where the spectral variation of the sky is of paramount importance, the frequency dependence
of a baseline's sampling of the $uv$-plane --- the same effect that gives rise to the delay transform
--- becomes a nuisance, since a baseline measuring different $uv$-modes as a function of 
frequency also measures different $k$-modes.  Furthermore, the frequency-dependent sampling of 
the $uv$-plane 
introduces spectral structure in nominally smooth foreground emission, leading to ``mode-mixing"
that complicates the separation of smooth-spectrum foreground emission from the
desired 21cm reionization signal.  One solution to this problem is to arrange antennas to create 
overlapping $uv$-coverage
at many frequencies, with different baselines providing samples of the same $uv$-mode at various frequencies.

In this paper, we propose an alternate approach that directly treats the frequency
dependent sampling pattern of a baseline.  We examine the per-baseline approach to measuring
21cm reionization introduced in \citet{parsons_et_al2012}, explicitly relating the delay transform to
the cosmological line-of-sight transform that is of importance for measuring reionization.
We then use the delay transform as a lens to help us understand how the separation between smooth-spectrum
foreground emission and the 21cm reionization signal is affected by the spectral response of
an interferometer.  Specifically, we will show that the geometric limit on the maximum delay at
which celestial emission can enter an interferometer sets a strict limit on the degree to which
smooth-spectrum foreground emission can corrupt measurements of $P_{21}(k)$ at higher magnitude
$k$-modes.

In \S\ref{sec:qual_dly}, we present a qualitative description of the how the delay transform
relates to measurements of the 21cm EoR power spectrum.  The goal of this section is develop
an intuitive, geometric interpretation of the delay transform, and to introduce terminology
that will recur throughout the paper.  In \S\ref{sec:quant_dly}, we present a mathematical
formalism for the delay transform. We use this formalism to explicitly address how the delay 
transform measures the 21cm EoR signal (\S\ref{sec:los_wavemodes}) and foreground emission 
(\S\ref{sec:map_foreground}).

\subsection{A Qualitative Picture of the Delay Transform for 21cm Experiments}
\label{sec:qual_dly}

The delay transform maps flat-spectrum emission from the sky to Dirac delta functions
in delay space.  Since different regions on the sky see different
physical geometric delays between the two antennas of a baseline, the delay transform serves as
a form of 1D, per-baseline ``imaging," as illustrated in Figure \ref{fig:spec_dly_shift}.  
Importantly, there is a maximum geometric delay, set by the length of the baseline, above which
no signal from the sky can enter.  For a zenith-phased array, this maximum delay occurs
at the horizon, so we refer to it as the ``horizon limit."  This fact bears repeating,
as it is crucial to the geometric understanding of the delay transform: flat-spectrum emission from the sky
cannot enter an interferometer baseline with a delay longer than that set by the length of 
the baseline.  

\begin{figure}\centering
    \includegraphics[width=.95\columnwidth]{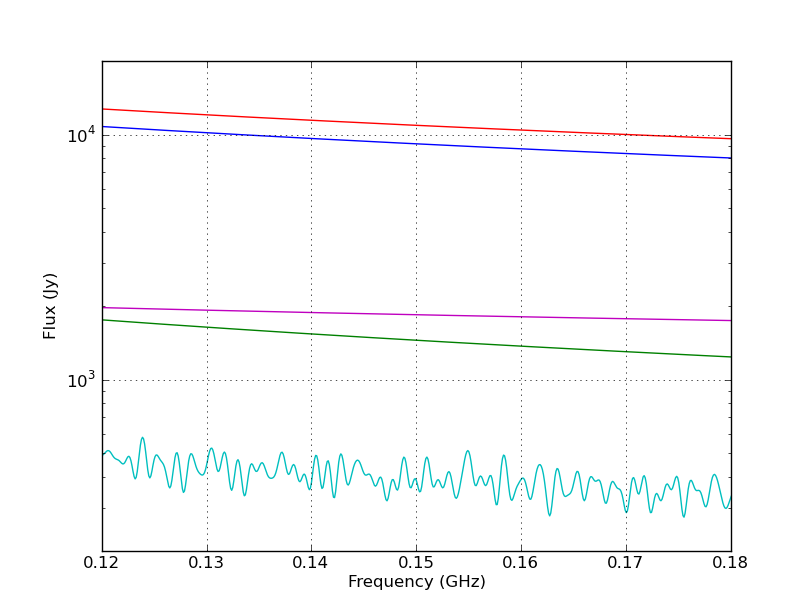}
\caption{
The spectra of five sources at random positions on the sky that were used to generate simulated
visibilities from which the delay spectra in Figure \ref{fig:dly_wfall} were calculated, using a model of 
PAPER's primary beam response.  All but one of the sources (cyan) have power-law
spectra versus frequency.  
}\label{fig:src_spectra}
\end{figure}

\begin{figure}\centering
    \includegraphics[width=.95\columnwidth]{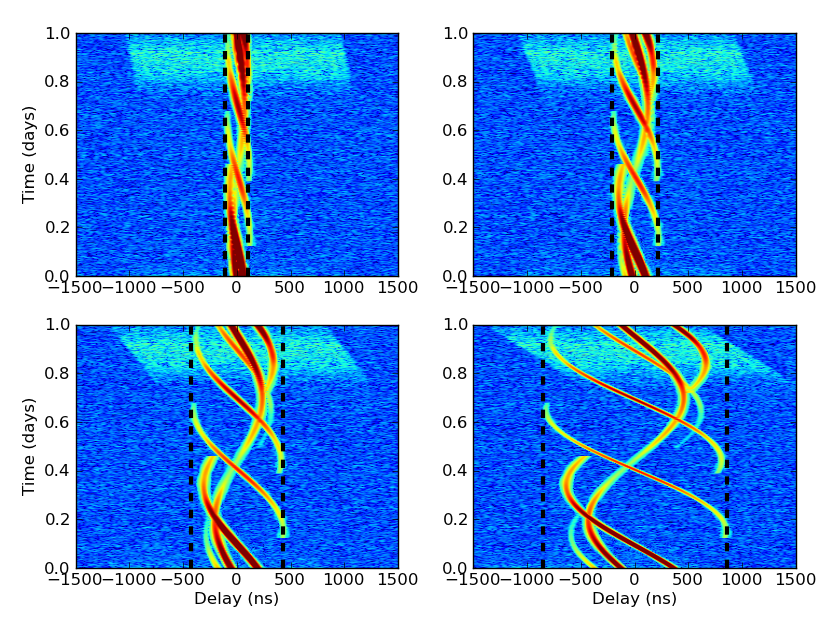}
\caption{
The delay spectra measured by baselines of four different lengths for a simulated sky consisting of several 
celestial sources, whose spectra are shown in Figure \ref{fig:src_spectra}.
The upper-left, upper-right, lower-left, and lower-right plots show 
delay spectra obtained by Fourier transforming a 60-MHz band centered at 150
MHz with a Blackman-Harris windowing function \citep{harris1978}
for east-west baselines of length 32, 64, 128, and 256 meters (16, 32, 64, and 128 wavelengths at 150 MHz), respectively.
Color scale denotes log$_{10}$-Jy amplitude, ranging from 1 (blue) to 5 (red).  
As in Figure \ref{fig:spec_dly_shift}, emission from sources with power-law 
spectra remains confined within the horizon limits (dashed vertical lines), while emission from the
source with an unsmooth spectrum (top region of each panel, corresponding to the source plotted in
cyan in Figure \ref{fig:src_spectra}) extends beyond these limits.
}\label{fig:dly_wfall}
\end{figure}

However, emission on the sky is not generally flat-spectrum, so
delay-domain ``images'' are smeared by a convolving kernel that reflects the
frequency dependence of celestial emission, instrumental responses, and the
finite bandwidth used in the delay transform.
Therefore, a Dirac delta function in delay-space, corresponding to emission from a specific
location on the sky, is broadened by convolution with this kernel.  Since the kernel corresponds to
the Fourier transform of any frequency structure in the spectrum of that source (whether intrinsic
or instrumental), emission with more frequency structure will have a broader signature in
delay-space.  In the limit of perfectly flat-spectrum emission, the response will return to a Dirac
delta function.  If the kernel is broad enough, however, a source can appear with non-negligible
power at delays beyond the horizon limit.  This statement forms the core of the delay
spectrum approach for isolating smooth-spectrum foregrounds from the 21cm EoR signal.  Due to
its intrinsically smooth spectrum, foreground emission will have a narrow convolving kernel, and so
will have rapidly decreasing power beyond the maximum delay of the horizon limit.  Source emission that is
not spectrally smooth, however, will create a very broad kernel in delay space, and so
will scatter power to delays well beyond the horizon limit, regardless of that source's actual
position on the sky.  This is illustrated in Figure \ref{fig:dly_wfall}, which shows the delay
spectra of 5 simulated sources, whose spectra are shown in Figure \ref{fig:src_spectra}.  Only the source with an unsmooth spectrum exhibits power outside
the horizon limit.  With the delay spectrum approach, one is, in a sense, looking for the
``sidelobes" of the EoR signal in delay-space that scatter power to high delays.  At these delays,
the dominant source of emission will not be foregrounds, but the EoR signal itself.

There are many issues that need to be addressed with a more quantitative treatment of the
above.  In \S\ref{sec:quant_dly}, we use a more rigorous framework to describe the delay transform.
In \S\ref{sec:los_wavemodes}, we use this formalism to relate delay modes to $k$-space.  
This entails explicitly handling the
frequency dependence of a baseline's length that gives rise to the delay transform, allowing us to derive
the point-spread function (PSF) of a delay mode in $k$-space that prevents a
one-to-one mapping of delay modes to line-of-sight $k$-modes. 
In \S\ref{sec:map_foreground},
we quantify the breadth of foreground emission in delay-space and the degree to which sidelobes of
foreground emission yield power beyond the horizon limit.

\subsection{Mathematical Formalism of the Delay Transform}
\label{sec:quant_dly}

\subsubsection{Notation and Coordinates}

To begin, let us define the various coordinates that we will use throughout the remainder
of the paper.  A baseline of an interferometer observes spatial Fourier modes of the sky as a
function of frequency, which we label with coordinates $(u,v,\nu)$.  Although these coordinates
naturally express an interferometric observation, 
for 21cm emission, they represent a mix between Fourier and real-space coordinates:
$u$ and $v$ are Fourier components of the transverse direction in the plane of the sky, and $\nu$ relates to the 
real-space, line-of-sight direction.
We define $\eta$ to be the Fourier transform of $\nu$, so that the coordinates $(u,v,\eta)$
form a 3D orthogonal coordinate system.  For 21cm emission, $|\vec{u}| \equiv
\sqrt{u^2 + v^2}$ and $\eta$ are related to the spatial Fourier modes $k_{\perp}$ and $k_{\parallel}$ 
(measured in $h\ {\rm Mpc}^{-1}$) via redshift-dependent constants of proportionality.  
For foreground emission, $|\vec{u}|$ and $\eta$ have no such cosmological interpretation.  As such,
$|\vec{u}|$ and $\eta$ are more general Fourier coordinates than $\vec k$.

\begin{figure}\centering
    \includegraphics[width=.75\columnwidth]{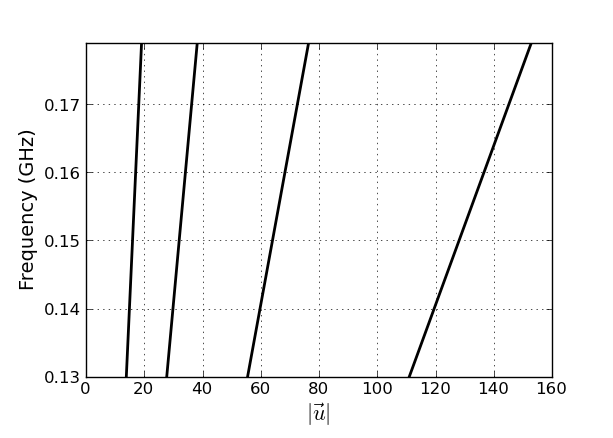}\caption{
Tracks sampled by baselines measuring
16, 32, 64, and 128 wavelengths (at 150 MHz), as a function of observing frequency,
in $(\vec u,\nu)$ space (solid, bottom axis).
The frequency dependence of the wavemode
sampled by an interferometer is one of the major complicating factors in foreground removal, and is also the reason that emission from celestial sources at different
positions on the sky maps to different regions of delay space.
The delay transform described in \S\ref{sec:dly_transform}
extracts Fourier modes measured along these tracks, rather than strictly along the frequency axis.
}\label{fig:kperp_vs_freq}
\end{figure}

The frequency dependence of a baseline's length (measured in
wavelengths) dictates that a baseline samples a sloped line through a $(u,v,\nu)$ cube.
This effect is illustrated for baselines of several lengths in Figure
\ref{fig:kperp_vs_freq}.  We define $\hat\kappa_b$ to be a unit vector pointing along the line
sampled by baseline $b$.
The delay transform, which is the Fourier transform of a single baseline's
visibility spectrum, is therefore a Fourier transform along the $\hat\kappa_b$ direction.
We refer to the Fourier complement of $\hat\kappa_b$ as the $\hat\tau_b$ direction, and single value of a 
delay spectrum as a $\tau$-mode.

Using this notation, the key point above is that since $\hat\kappa_b$ is not
parallel to $\hat\nu$ (as indicated in Figure \ref{fig:kperp_vs_freq}), $\hat\tau_b$ will not be 
parallel to $\hat\eta$.  This means that $\tau$-modes do not directly correspond to
cosmological $k_{\parallel}$-modes.  The ramifications of this fact are discussed in 
\S\ref{sec:los_wavemodes}, where we explicitly compute the PSF of $(|\vec{u}|, \tau)$-modes
in $(k_{\perp},k_{\parallel})$-space.  

\subsubsection{The Delay Transform}

The visibility response, $V$, of a single baseline,
assuming that all non-geometric components of visibility phase have been calibrated and removed,
can be expressed by:
\begin{equation}
    V(u,v)=\int{\frac{dl~dm}{\sqrt{1-l^2-m^2}}~A(l,m)I(l,m)}{e^{-2\pi i\left(u \, l+v \, m+w (\sqrt{1-l^2-m^2} - 1)\right)}},
    \label{eq:w_visibility}
\end{equation}
where $l\equiv\sin\theta_x$ and $m\equiv\sin\theta_y$ relate to angular coordinates in the plane of the sky,
$A(l,m)$ is a windowing function describing the spatial and spectral response of an
interferometric pair of antennas, and $I(l,m)$ is the specific intensity.  Note that this
equation is implicitly frequency-dependent, both in $A(l,m)$ and $I(l,m)$, but also in the 
Fourier components $u$,$v$, and $w$.  As shown in PB09, we can re-express equation \ref{eq:w_visibility} 
for a single baseline as:
\begin{equation}
    V_b(\nu)=\int{dl~dm~{A(l,m,\nu) I(l,m,\nu)e^{-2\pi i\nu\tau_g}}},
    \label{eq:meas_eq}
\end{equation}
where $\nu$ is spectral frequency, $b$ is used to indicate this delay spectrum
comes from a single baseline, $\vec b$,
 and
\begin{equation}
    \tau_g\equiv\frac{\vec b\cdot\hat s}{c} = \frac1c\left(b_xl+b_ym+b_z\sqrt{1-l^2-m^2}\right)
    \label{eq:dly}
\end{equation}
is the geometric group delay associated with the projection of the baseline vector 
$\vec b\equiv(b_x,b_y,b_z)$ toward
the direction $\hat s\equiv(l,m,\sqrt{1-l^2-m^2})$, as illustrated in Figures \ref{fig:spec_dly_shift} and
\ref{fig:dly_wfall}.  The visibility-domain coordinates 
$\vec u\equiv(u,v,w)$ are related to baseline coordinates by
$\vec u=\nu\vec b/c$.

The Fourier transform of the visibility spectrum from one
baseline (as written in equation \ref{eq:meas_eq}) along the frequency axis yields the delay transform:
\begin{equation}
    \tilde{V}_b(\tau)=\int{dl~dm~d\nu~{A(l,m,\nu) I(l,m,\nu)e^{-2\pi i\nu(\tau_g-\tau)}}}.
    \label{eq:dtransform}
\end{equation}
In constrast,
if we extend the definition of the visibility from equation \ref{eq:w_visibility}, as done in P12, to include 
a Fourier transform along the frequency axis (and ignore the $b_z$ component of a baseline; 
we discuss issues arising
from this simplification in \S\ref{sec:gridding}), we have
\begin{equation}
    {\tilde V}(u,v,\eta)=\int{dl~dm~d\nu~A(l,m,\nu)I(l,m,\nu)}{e^{-2\pi i(u \, l+v \, m+\eta \,\nu)}},
    \label{eq:vhat_definition}
\end{equation}
where $\eta$ is, as previously described, the Fourier transform of frequency $\nu$, and is orthogonal
to $u$ and $v$.
We have now explicitly noted the frequency dependence of $A(l,m)$ and $I(l,m)$\footnote{In P12, 
it was noted that this definition ignores the frequency dependence of $(u,v)$, which
vary by as much as 6\% over an 8-MHz bandwidth.  It was briefly argued that,
given the configuration of most EoR experiments, the $\vec k_\perp$ component of $\vec k$
that arises from $(u,v)$ contributes negligibly to $|\vec k|$, so that this approximation does not substantially
affect sampling of $P_{21}(\vec k)$, nor does it change the derived sensitivities.  The purpose of this present work is to explore the effects of this frequency dependence in greater detail.}.
Since $u,v$ and $\eta$ are directly related to $k$-modes of a cosmological volume, $\tilde{V}$ is
a direct probe of the 21cm power spectrum.  However, as we have stated, the transform in 
equation \ref{eq:vhat_definition}
cannot be calculated from a single baseline's visibility spectrum, due to the frequency dependence
of $u$ and $v$ sampled by one baseline.

We wish to explore the effects of using $\tilde{V}_b(\tau)$ (equation 
\ref{eq:dtransform}) from 
one baseline as a substitute for equation $\tilde{V}(u,v,\eta)$ (equation \ref{eq:vhat_definition}), 
which requires multiple baselines of data.  Put another way, we ask: what is the effect
of using delay-modes to measure $k$-modes?  The result will be that $\tilde V(\tau)$ has a 
PSF that mixes Fourier modes along the $\hat\eta$ direction.  In order to derive the
effects of mode-mixing, we will directly examine the result of taking a Fourier
transform along the direction inherently sampled by a baseline---a direction we have called $\hat\kappa_b$.
We can now express this direction as a unit vector given by:
\begin{equation}
    \hat\kappa_b\equiv\frac{\nu\hat\nu + \frac{b_x\nu}{c}\hat u + \frac{b_y\nu}{c}\hat v}{\sqrt{\nu^2+(b_x\nu/c)^2+(b_y\nu/c)^2}}.
\end{equation}
Similarly, the $\hat\tau_b$ direction that is the Fourier complement of $\hat\kappa_b$ 
can also be expressed in terms of previously defined coordinates:
\begin{equation}
    \hat\tau_b = \frac{\frac1\nu\hat\eta + \frac{c}{b_x\nu}\hat l + \frac{c}{b_y\nu}\hat m}{\sqrt{(1/\nu)^2+(c/b_x\nu)^2+(c/b_y\nu)^2}}.
\end{equation}
The $\hat\tau_b$ direction
deviates from the $\hat\eta$ direction with a component in the image plane that depends on baseline 
length.  This component causes a $\tau$-mode to sample different $\eta$-modes
at different $(l,m)$-coordinates.
It is the projection of $\hat\tau_b$ along the angular sky coordinates that gives
rise to the delay transform's mapping of different points on the sky to 
different geometric delays.

\subsection{The PSF of $\tau$-modes in $k$-space}
\label{sec:los_wavemodes}

We now move to
determining the PSF of delay-modes in $k$-space to examine the degree to which they may be regarded
as samples of the three-dimensional power spectrum of reionization.
A single delay-mode, $\tilde{V}_b(\tau_0)$, has an extended shape in both the $k_\perp$ and
$k_\parallel$ directions.
We begin with the PSF of a $\tau$-mode in the $k_\perp$ direction, or equivalently, in the $uv$-plane.
This PSF, which we will call $W_{\tau,b}(u,v)$, reflects the inherent width of the antenna beam response in the 
$uv$-plane, $\tilde A(u,v,\nu)$, convolved by 
the frequency dependence of the $uv$-coordinate sampled by a baseline $b$:
\begin{equation}
    W_{\tau,b}(u,v)\equiv\int_{\nu_0-B/2}^{\nu_0+B/2}{d\nu~\tilde A\left(u-\frac{b_x\nu}{c},v-\frac{b_y\nu}{c},\nu\right)~e^{-2\pi i(\nu+b_x\nu/c+b_y\nu/c)\tau}},
\label{eq:w_tau_uv}
\end{equation}
where $\nu_0$ and $B$ are the center and width of the frequency band used in the delay transform, respectively.
Over the largest ($\sim\!8$-MHz) bandwidths that fit within the evolutionary timescales of the 21cm EoR signal
\citep{wyithe_loeb2004,furlanetto_et_al2006}, the $uv$-modes sampled by $V(\vec\kappa)$ are localized in the $uv$-plane.
For baseline lengths less than 300 m and for $k_\parallel\gg0.01\ h\ {\rm Mpc^{-1}}$, 
the 6\% maximum variation in $(u,v)$ over
this bandwidth is much smaller than the scales over which $\Delta_{21}^2(\vec k)$ evolves. Hence, the PSF
broadening in the $\vec k_\perp$ direction that results from the slight difference between 
$uv$-coordinates
sampled at the upper and lower edges of the band
has a negligible effect on the inferred value of $\Delta_{21}^2(\vec k)$.
The Fourier transformation of $V(\vec\kappa)$ to $\tilde V(\vec\tau)$ does not affect which 
$uv$-modes were inherently sampled, and hence the PSF of $\tilde V(\vec\tau)$ remains peaked in the $\vec k_\perp$
direction.

\begin{figure}\centering
    \includegraphics[width=.75\columnwidth]{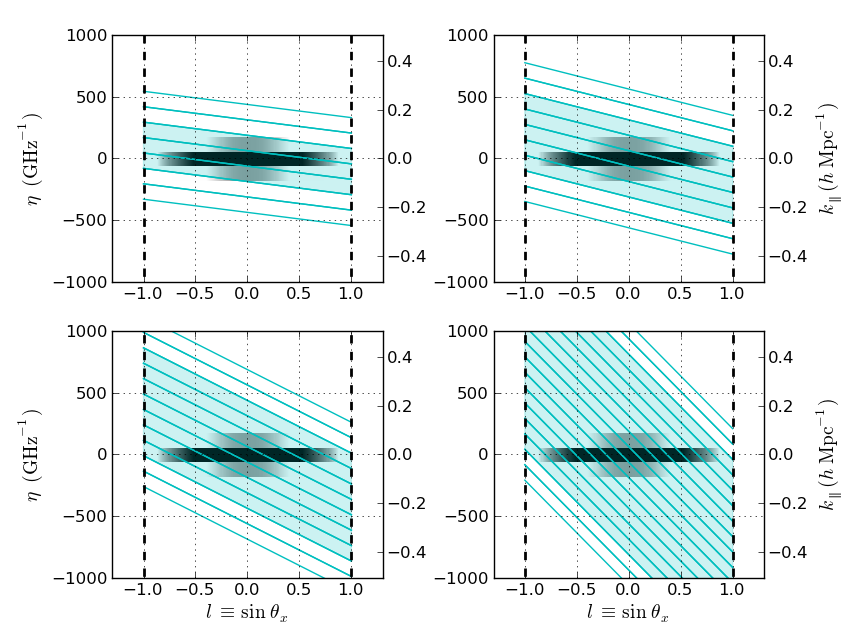}\caption{
The response in $(l,\eta)$ coordinates of delay bins measured by baselines of four different lengths, all
chosen, without loss of generality, to be in the east-west direction.
The grayscale shading in the center of each plot shows the response of the PAPER primary beam, 
integrated over the direction perpendicular to the baseline (i.e., in the $m$ direction, along the delay 
bins indicated in Figure \ref{fig:lm_iso_delay_contours}), and Fourier transformed in frequency over an 8-MHz band
(centered at 150 MHz) at each
$l$ coordinate to yield $\int{\tilde A(l,m,\eta)~dm}$.  Grayscale indicates response relative to peak amplitude
in $\log_{10}{\rm mK}^2$ units, ranging from 0 (black) to -2 (white).  In this space, the beam 
spans a broad range of $l$ (it has a wide FWHM) but is fairly narrow in $\eta$ (it is very
smooth in frequency).  
Overlaid on
the beam representation are the regions of sensitivity of delay-spectrum bins for baselines of length
32 (upper-left), 64 (upper-right), 128 (lower-left), and 256 (lower-right) meters, corresponding to 16, 32,
64, and 128 wavelengths, respectively.  The area between two colored
lines corresponds to the region to which a single $\tau$-mode of that baseline's delay spectrum is sensitive.  
As in Figure \ref{fig:spec_dly_shift}, the
thick dashed line denotes the horizon cutoff. 
Each $\tau$-mode probes a range of $\eta$-values across the field-of-view, 
although shorter baseline lengths reduce the degree to which delay-bins mix $\eta$-modes.  
To aid the eye, we have shaded $\tau$-bins that intersect the primary beam response at greater than 1\% of peak
amplitude, indicating delay modes where instrumental response scatters perfectly smooth-spectrum celestial emission.
For each baseline length, there also exist unshaded delay bins
where significant power can only exist if celestial emission itself has structure
at sufficiently high $\eta$ (i.e., it has rapidly fluctuating frequency structure).  
Since foreground emission at these frequencies is predominantly smooth-spectrum 
\citep{furlanetto_et_al2006},
this approach allows the dominant foregrounds to the 21cm EoR signal to be avoided by employing antennas
with smooth primary beams versus frequency (to constrain the grayscale width in the $\hat\eta$ direction) and
by employing shorter baselines that minimize the mixing of $\eta$-modes within delay bins (by flattening the
slope of the colored lines).
}\label{fig:dly_eta_sky_resp}
\end{figure}

Examining the PSF of $\tau$-modes in the $k_\parallel$ direction requires greater attention.  At first blush,
one might expect the breadth of $\tau$-modes in the $(l,m)$-directions (see Figure
\ref{fig:dly_eta_sky_resp}) to cover such a broad range of $\eta$-modes 
that the PSF of $\tilde V(\tau)$ in $k$-space would be irretrievably compromised.  On closer
examination, however, we see that
the extent of any $\tau$-mode in $(l,m)$-domain is controlled by the angular size of the baseline's 
primary beam ($A(l,m,\nu)$ in equation \ref{eq:meas_eq}), which
is fundamentally limited to be zero outside of $-1\le l,m\le1$ by the horizon.
As illustrated in Figure \ref{fig:dly_eta_sky_resp}, any single delay bin
would probe (i.e. mix) all $\eta$-modes, were it not for the hard boundary of the horizon limit.
This statement is another way of expressing the key fact that the horizon sets a fundamental maximum delay for signals on the sky.

\begin{figure}\centering
    \includegraphics[width=.75\columnwidth]{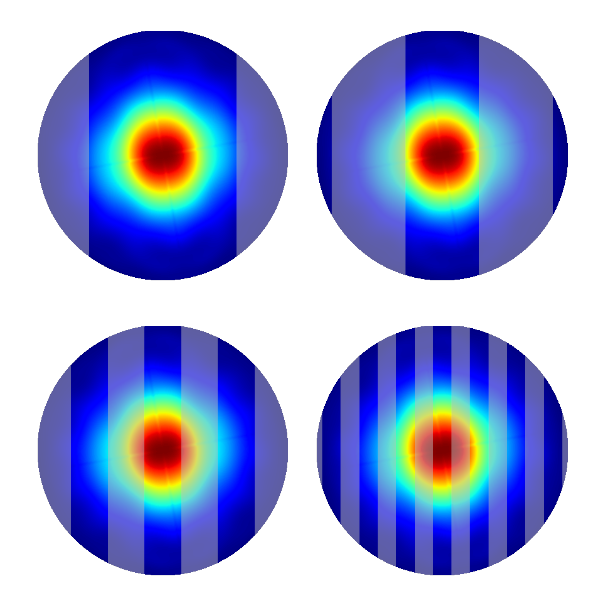}
\caption{
The delay bins (alternating shaded/unshaded regions) resulting from the delay 
transform applied to an 8-MHz band of data from east-west oriented baselines
of length 32 (upper-left), 64 (upper-right), 128 (lower-left), and 256 (lower-right) meters, 
corresponding to 16, 32, 64, and 128 wavelengths, respectively.
Bin responses are shown projected over the spatial response of the 
PAPER primary beam at 150 MHz.  Color scale illustrates normalized power response, ranging from 0 (blue) to
1 (red) at zenith.  Note that the number of delay bins shown overlapping the field-of-view in this
figure corresponds to the number of shaded bins in Figure \ref{fig:dly_eta_sky_resp} that indicate regions
where perfectly smooth-spectrum celestial emission can enter a delay mode as a result of instrumental response.
}\label{fig:lm_iso_delay_contours}
\end{figure}

As a result, we may consider
the width of a $\tau$-mode in the $\hat\eta$ direction to be the inherent $1/B$ width 
set by the bandwidth $B$
used in the Fourier transform, convolved by the PSF of $\tilde A(l,m,\eta)$ in delay coordinates.  This
PSF, which we will call $W_{\tau,b}(\eta)$, can be computed explicitly by 
integrating $\tilde A(l,m,\eta)$ along iso-delay contours (see
Figure \ref{fig:lm_iso_delay_contours}):
\begin{equation}
    W_{\tau,b}(\eta)\equiv\int{dl~dm~\tilde A(l,m,\eta)~\delta_D\left(\frac{b_xl}{c}+\frac{b_ym}{c}+\eta-\tau\right)},
\label{eq:w_tau_eta}
\end{equation}
where $\delta_D$ is the Dirac delta function.  Convolving by $W_{\tau,b}$ captures how
celestial emission that is uniformly distributed across the field-of-view with a characteristic spectrum 
is measured by a baseline of fixed physical length; the measured delay spectrum represents an integral
over the sky of the
inherent delay spectrum of the emission entering at a different delay for each point
on the sky, multiplied by the primary beam response in that direction.  Hence, as shown
in Figure \ref{fig:dly_eta_sky_resp}, $W_{\tau,b}(\eta)$ accounts for the fact that the $\eta$-mode sampled by any 
chosen delay bin changes linearly across the
sky with a slope that depends on the baseline length.

Because $W_{\tau,b}$ captures how baseline length and the spatial and spectral variation of the primary beam
modifies the inherent delay spectrum of foreground emission, it is the ultimate metric for judging which baseline
lengths and primary-beam response patterns avoid scattering smooth-spectrum foreground emission into
$k$-modes that are of interest for measuring 21cm reionization.  The two components of $W_{\tau,b}$ ---
the Fourier transform of the primary beam along the frequency direction $\tilde A(l,m,\eta)$ illustrated
in grayscale in Figure \ref{fig:dly_eta_sky_resp}, and the changing slope of delay-bin responses
in $(l,m,\eta)$-space illustrated with colored lines --- together determine which delay-bins are corrupted
by foregrounds (the shaded regions in Figure \ref{fig:dly_eta_sky_resp}).  By limiting the slope of delay-bin
responses by using shorter baseline lengths, by designing antennas with spectrally-smooth primary 
beam responses to limit their extent in the $\hat\eta$ direction, and (somewhat less critically) by 
limiting the spatial breadth of the primary beam response in the $(l,m)$ direction, it is possible to
reduce the width of $W_{\tau,b}$, and thereby reduce the level of foreground corruption in
$k$-modes of interest.  It is worth noting that in order to precisely determine what amplitudes of $W_{\tau,b}$ are
acceptable, future work will be needed to determine the inherent brightness and spectral smoothness of 
foregrounds.

\begin{figure}\centering
    \includegraphics[width=.75\columnwidth]{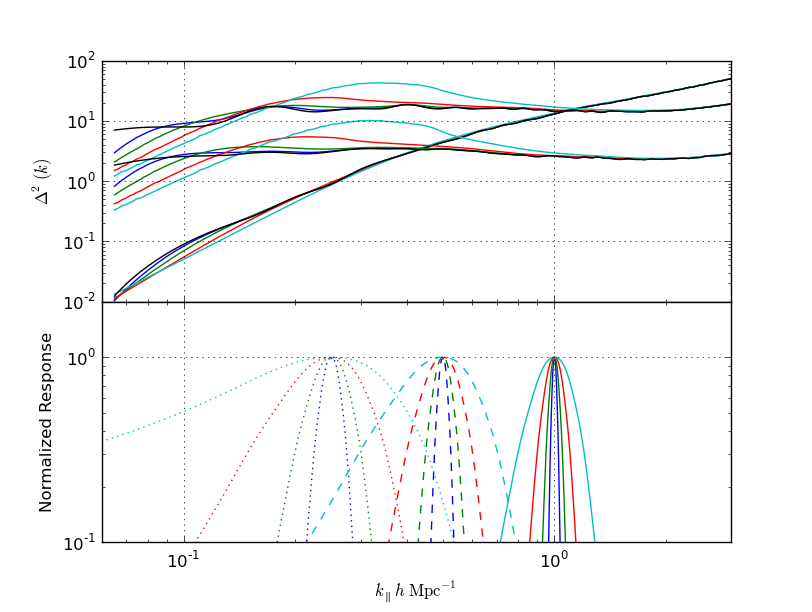}\caption{
Top: the effect of the delay-transform PSF on measured power spectra of 21cm reionization, using
model power spectra at various stages of reionization from \citet{lidz_et_al2008}.
Bottom: the $k_\parallel$ response ($W_{\tau,b}(\eta)$, from Equation \ref{eq:w_tau_eta}), 
of a delay-transform bin arising from
the sloped response in $k_\parallel$ shown in Figure \ref{fig:dly_eta_sky_resp}, integrated over the
primary beam response of a PAPER dipole,
at bins centered at $k_\parallel=0.25$, $0.5$, and $1.0\ h\ {\rm Mpc}^{-1}$ (dotted, dashed, and sold plots, respectively).  
Color indicates baseline
length in wavelengths at 150 MHz of 16 (blue), 32 (green), 64 (red), and 128 (cyan).
The width of the response of a delay-transform bin
arises from the inherent scaling of $uv$-coverage with frequency, or equivalently, from the
width of the primary beam in delay-space.  Only for the longest baselines does mode-mixing
substantially affect measurements of the power spectrum of high-redshift 21cm emission.
    }\label{fig:pk_resp_of_dly_bin}
\end{figure}

If, for a moment, we take $A(l,m,\eta)$ to be unity across the entire sky, we may set an upper bound on the width of 
a $\tau$-mode in the $k_\parallel$ direction by using the slope 
of a baseline's response in $(l,m,\eta)$-space,
imposing the horizon limits on $(l,m)$, and then converting $\eta$ to $k_\parallel$ using a
cosmological scalar, $Y(z) = dk_{\parallel}/d\eta$, that
relates a frequency interval to a comoving physical size (see \citealt{furlanetto_et_al2006,parsons_et_al2012} for computations of $Y$).
Without loss of generality, we adopt a strictly east-west
baseline orientation such that $m$ can be neglected:
\begin{equation}
    \Delta k_\parallel\le\Delta l\frac{dk_\parallel}{dl}
                      =\Delta l\frac{dk_\parallel}{d\eta}\frac{d\eta}{dl}
                      =\Delta l~Y\frac{|\vec b|}{c}.
\end{equation}
For a baseline length of $|\vec u|=20$ at 150 MHz ($z=8.5$), we get $\Delta k_\parallel\approx0.13~h{\rm Mpc}^{-1}$ using the most conservative choice of
$\Delta l=2$, for emission entering from the entire sky;
for a narrower primary beam, $\Delta l$ can be somewhat smaller.
The final step uses $d\eta/dl = |\vec{b}|/c$, and comes from the fact that the length of a baseline
sets the slope of that baseline's $uv$-sampling pattern versus frequency.
For comparison, the width of $W_{\tau,b}$ in the $uv$-plane for PAPER antennas over an 8-MHz bandwidth
is approximately 50 times narrower 
($\Delta k_\perp\sim0.002 h\ {\rm Mpc}^{-1}$),
confirming that the width of $W_{\tau,b}$ is predominantly in the $\hat\eta$ direction.
Figure \ref{fig:pk_resp_of_dly_bin} shows $W_{\tau,b}(\eta)$, the PSFs in the $\hat\eta$ direction, for several baseline lengths, computed explicitly using the spatial and
spectral primary beam
response of PAPER elements.
Since $\Delta_{21}^2(k)$ is expected to evolve on ${\rm log}~k$ scales \citep{mcquinn07,trac07},
the width of $\tau$-modes in the $k_\parallel$-direction only becomes important on scales 
of $\Delta k\sim k$.
It is worth noting that although
adjacent $\tau$-bins may appear to have overlapping response in $k$-space,
the modes sampled are actually statistically independent, as shown by their 
orthogonality in $(l,m,\eta)$-space in Figure \ref{fig:dly_eta_sky_resp}.

% DON'T THINK IT IS NECESSARY TO MAKE ARGUMENT THAT FOR EOR, INTEGRATING POWER LAW IN $k$ OVER FOV YIELDS POWER LAW
% OF SAME SLOPE AT SIMILAR AMPLITUDE.

\subsection{Mapping Foregrounds to $k$-space}
\label{sec:map_foreground}

In the previous section, we reiterated how, for a substantial region of $k$-space, $\tau$-modes may be considered
direct measures of $P_{21}(\vec k)$, owing to their peaked response in $k$-space \citep{parsons_et_al2012}.  
Using a delay-spectrum methodology, we showed how the PSF of $\tau$-modes in $k$-space can be computed
explicitly, and we set upper bounds on the breadth of the PSF by invoking geometric limits on
the delay at which celestial emission can enter an interferometer.
However, the most powerful aspect
of the per-baseline, delay-spectrum approach to measuring 21cm EoR concerns the isolation of smooth-spectrum foreground emission described in \S\ref{sec:qual_dly}.  
In this section, we use the formalism of \S\ref{sec:quant_dly} to identify a threshold
in delay-space beyond which $\tau$-modes are uncorrupted by smooth-spectrum foreground emission, making
them effective probes of $\Delta_{21}^2(k)$.

To begin, let us consider foreground emission consisting of a discrete set of point sources.
Neglecting non-geometric
delay terms, we translate the 
frequency-domain visibility response to delay-domain for foreground emission consisting of a discrete set of
point sources:
\begin{equation}
  \begin{aligned}
    \tilde V_b(\tau)&=\int_{-\infty}^{\infty}{
      \left[\sum_n{
        A(\nu,\hat s_n) S_n(\nu)
        e^{-2\pi\nu \tau_{n}}
      }\right]
        e^{2\pi i\nu\tau}\ d\nu
    }\\
    &= \sum_n{\left[
        \tilde A(\tau,\hat s_n)
        \ast
        \tilde S_n(\tau)
        \ast
        \delta_D\left(\tau_{n}-\tau\right)
    \right]}
    \label{eq:dly_t}.
  \end{aligned}
\end{equation}
Here, we see a measured visibility expressed as a discrete sum over point sources, each entering at
a different delay with a different inherent frequency spectrum.  The delay transform maps flux from each
celestial source to a Dirac delta function, $\delta_D$, centered at the corresponding group delay, 
convolved by a kernel representing the Fourier transforms of
frequency-dependent
interferometer gains, $\tilde A(\tau,\hat s_n)$, and the inherent spectrum of each source, $\tilde S_n$.
We should note that the
reason for examining a point-source foreground is entirely pedagogical and does not reflect a loss of
generality; we could just as well have considered an integral over
$S(l,m,\nu)$ and have computed the delay corresponding to each $(l,m)$ coordinate explicitly.

As derived in the previous section, there is a geometric maximum geometric delay,
$\tau_{max}$, at which flux can enter an interferometer.
For a primary beam response such as PAPER's that covers the
entire sky, $\tau_{max}$ is bounded by the horizon.  Beyond this maximum
delay, any observed flux must be the result of sidelobes of the $\tilde A\ast\tilde
S_n$ kernel convolving $\delta_D(\tau-\tau_n)$ for some
$-\tau_{max}\le\tau_n\le\tau_{max}$, as shown in Figure \ref{fig:dly_wfall}.
Ignoring the antenna primary beam response $\tilde A$ for a moment,
we can see that for sources with sufficiently smooth spectra, $\tilde S(\tau)$ will
be quite narrow, and therefore, their contribution to foreground emission
will be narrowly confined around $\tau_n$.  For emission
that is unsmooth versus frequency, such as the expected 21cm reionization signal, $\tilde S(\tau)$ scatters power 
well beyond $\tau_{max}$, and as shown in \S\ref{sec:los_wavemodes}, these delay-modes
beyond the horizon represent samples of $P_{21}(k)$.

The key to managing foregrounds using the delay-spectrum technique
is for both foreground emission and instrumental responses
to be sufficiently smooth in the $\hat\kappa_b$ direction that foreground emission remains
tightly bound around the maximum delays in delay-space.  The width of $\tilde A*\tilde S$
in delay-space, along with the baseline length that determines $\tau_{max}$,
fundamentally determines which $k_\parallel$-modes are corrupted by foregrounds.
Fortunately, the dominant foregrounds to the 21cm reionization signal are
expected to be spectrally smooth, with the possible exception of polarized
galactic synchrotron emission, whose treatment we will discuss briefly in
\S\ref{sec:discussion}, but will largely defer to a future paper.  In \S\ref{sec:simulation}, we will
examine the foregrounds in greater detail in the context of the delay transform
using simulations.

Besides baseline length (where, barring instrumental systematics, shorter is
better\footnote{Interestingly, the delay-spectrum foreground isolation approach can also operate
effectively on auto-correlations to reveal unsmooth spectral features.  
Step-functions in auto-correlation spectra caused by the global 21cm reionization
signature have been ruled out \citep{bowman_rogers2010}, and current models favor the prolonged
evolution of the global 21cm signal.  However, a fact that has been widely overlooked
is that with adequate mastery of systematics and foreground removal, auto-correlations can 
also constrain $\Delta_{21}^2(k)$.  This is possible because the delay-spectrum
of an auto-correlation yields samples of $k$-modes where $k_\perp$ is 0, but $k_\parallel$
can still access $k$-modes of interest.  The closest acknowledgement of this prospect in
the literature comes
from \citet{bittner_loeb2011}, where they discuss the possibility of sharp spectral features in
auto-correlation spectra arising from fluctuations in the average temperature and ionization
fraction as a function of redshift.  The sensitivity requirements for
sampling $\Delta_{21}^2(k)$ via auto-correlations are derived from the same sensitivity calculations
as for cross-correlations in P12.  While the systematics of auto-correlations can be 
more difficult to tame than for cross-correlations, this approach to detecting reionization is an
interesting one.}), the
instrument design parameter that is most important to the success of the delay transform is the
smoothness of $A(l,m,\nu)$.  Furthermore, as shown in Figure \ref{fig:lm_iso_delay_contours}, delay-modes have projections along $l,m$, and
$\nu$, making it imperative that $A(l,m,\nu)$ be both spatially and spectrally smooth, since spatial and spectral responses of an interferometer are commingled in the delay-spectrum
approach.  To avoid scattering smooth-spectrum foregrounds
into delay bins beyond the geometric horizon limit, the spatial structure of the primary
beam cannot evolve rapidly over the frequency bands used in the delay transform.  Frequency-dependent sidelobes can
be particularly problematic, especially for larger dishes where the change in dish diameter over an 8-MHz bandwidth
can be greater than a wavelength.

A traditional antenna
metric that particularly succinctly captures these constraints is the frequency dependence
of the standing-wave ratio (SWR, see, e.g., Ch. 10-6 of \citealt{kraus_carver1973}).
The amplitude of spectral features in the SWR for the antenna signal path
must be kept sufficiently small such that, when the antenna response multiplies
the power density of correlated foreground emission between two antennas, the leakage into
delay-modes of interest is much smaller than the expected 21cm reionization signal.  Coarsely, the
most straightforward way to achieve this is by limiting standing waves in the system to scales 
larger than the bandwidth used in the delay transform.  This can be done by
limiting the geometric size of the antenna element so
that the light crossing time of the aperture is much smaller than the delay scales of interest, and by
ensuring proper termination of long transmission lines.  If all other smoothness constraints
are met, it can also be advantageous to
limit the angular size of the primary beam, which both improves sensitivity \citep{parsons_et_al2012}
and shortens the effective length of the baseline in delay space, making a
longer baseline behave as a shorter one, as given by the following relation between the maximum
delay $\tau_{max}$ associated with smooth foreground emission and the 
maximum angle from zenith $\theta_{max}$ at which a primary beam exhibits response:
\begin{equation}
\tau_{max}=|\vec b|\sin\theta_{max}/c.
\end{equation}
In practice, however, foregrounds are very bright, and designing antenna elements with
primary beams responses that attenuate foregrounds to below the 21cm reionization signal level 
outside of $\theta_{max}$ may not be achievable while maintaining the spectral smoothness constraint
mentioned above.

Another important instrument design choice that affects the efficacy of the delay
transform is the total correlated bandwidth of the interferometer.  Although the 
Fourier transform used for generating samples of $\Delta_{21}^2(\vec k)$ is bounded to $\sim\!8$ MHz
by the expected rapid evolution of the peak reionization signal versus redshift, the entire
effective bandwidth of the instrument can be used in the delay transform for the purpose of modeling
and removing foreground emission.  This is possible because smooth-spectrum foreground emission can be
coherently added over very large bandwidths. As the bandwidth used in the delay transform approaches
the observing frequency, the resolution in delay domain approaches the imaging resolution 
of synthesis imaging.  Enhanced delay-domain resolution can make it possible to 
model and remove foreground near to the horizon with finer control than is possible with the narrow
bandwidths used in the cosmological transform.  In addition, wider bandwidths improve foreground
sensitivity for the purpose of removing extrapolated foregrounds in bands of interest.  Finally,
as will be discussed in \S\ref{sec:implementation}, wide bandwidths improve the deconvolution process
that is used to compensate for poorly-sampled frequency channels with high RFI occupancy.

\section{Implementation Details}
\label{sec:implementation}

In this section, we examine some critical details in the implementation of an analysis
pipeline based around the delay-spectrum technique for foreground avoidance.  In particular,
we describe how data flagging can be accommodated without drastically compromising
foreground containment, and how measurements are binned and coherently added to improve sensitivity.

\subsection{Reducing sidelobes of foregrounds arising from frequency-domain flagging}
\label{sec:implementation:cleaning}

Extended sidelobes in delay-space can have a variety of causes beyond unsmooth intrinsic source spectra.  Of particular
concern are unsmooth spectral responses introduced by flagging RFI-contaminated
data. \citet{parsons_backer2009} highlighted how deconvolution algorithms
such as CLEAN \citep{hogbom1974} can effectively compensate for the
effects of flagged data.  Because foreground signals are coherent over wide
bandwidths, we apply the delay transform and deconvolution over the entire
observing band to produce the best estimate of the group delay and
spectral-shape convolution kernel associated with sources.  CLEAN components in
delay-space are restricted to lie within the maximum delays imposed
geometrically by the horizon.  We then subtract the wide-band deconvolved
foreground model from the measured visibilities to suppress the scattering of
foreground emission off of unsmooth sampling weights relating to RFI
flagging.
In essence, delay-space CLEAN is a direct analog of the image-domain 
modeling proposed for foreground removal \citep{paciga_et_al2011,datta_et_al2009}.  The difference here
is that it is acting per-baseline, and that, in addition to selecting CLEAN components
in delay bins that map to image-domain coordinates, delay-space CLEAN is also restricted
to models that produce smooth spectral responses.

Band edges also cause extended
sidelobes in delay-domain spectra.  To mitigate these effects in the work presented here, 
we use a 3-tap, 100-channel Polyphase Filter Bank \citep{vaidyanathan1990} with a Blackman-Harris window 
\citep{harris1978} to perform the delay-transform.
The frequency-domain $\sin x/x$ response of the PFB in this application trades some locality in the frequency domain in order 
to improve filter steepness in the D-domain.  This is acceptable because the PFB nonetheless strongly weights data 
from the band center in frequency domain, and the filter steepness in delay domain is critical for accessing
$k$-modes of interest.

\subsection{Gridding and Integration}
\label{sec:gridding}

In order to build sensitive measurements, it is necessary to coherently add measurements that sample
the same Fourier modes before samples of independent Fourier modes are squared and combined.
In order to represent samples of the same sky signal, baselines must sample the same $(u,v,w)$
coordinate at the same sidereal time.  Of course, there is a coherence interval in $(u,v,w,t)$-space,
whose width depends on instrumental parameters, where samples effectively add in-phase.  Within a frequency interval where the primary beam does not evolve substantially, the
width of the coherence interval in the $uv$-plane is determined by a PSF that is the Fourier transform
of the primary beam response in $(l,m)$ sky coordinates. This PSF can be used to determine optimal
weights for gridding measurements into $uv$-bins \citep{morales_matejek2009}.  Neglecting the time-dependence
of a baseline's $uv$-sampling (which is accounted for in gridding), the time interval over which
observations may be considered to be of the same sky relates to how long
the primary beam illuminates the same region of sky.  The width of the primary-beam response in
the east/west direction determines the weighting for adding observations from different sidereal times into
time bins.  For PAPER, these independent time bins are spaced approximately 2 hours apart.

Choosing a gridding interval along the $w$ direction is somewhat more involved.  While many arrays
are configured to be nearly isoplanar relative to a zenith pointing, it is nonetheless the case that within
a 2-hour observation (the binning time interval suggested for PAPER), baselines phased
to a zenith-transiting phase center will have an non-negligible and time-dependent projection along
the $w$ direction.  The phase offset associated with this $w$ term can be compensated for at phase
center by projecting observations at various $w$ coordinates down to the $w=0$ plane,
and hence, the same $w$ bin.  However,
as is well-studied in context of standard interferometric imaging, the spatial
dependence of this phase offset causes observations that are projected from different $w$ coordinates 
to decohere from one another as a function of distance
on the sky from phase center.  This problem is exacerbated by longer
baseline lengths.  As we will discuss below, we suggest that it is desirable to add entire spectra
together with uniform weighting versus frequency. This, unfortunately,
is not compatible with the W-projection technique that has come to be the standard method for compensating
for a spatially-dependent phase term \citep{cornwell_et_al_2005_347}.

The other option for maintaining uniform weighting across the band in the face of non-zero $w$ terms is
to define an interval over which spatial decoherence from $w$ has a negligible effect on the resulting
measurement, and then to use this interval to grid in $w$.  A reasonable scale for defining a grid
in $w$ is the point at which phase errors at the half-power point of the primary beam approach a radian.
For a PAPER beam radius of $\sim\!30^\circ$, this interval is approximately 1.2 wavelengths.  Unfortunately,
given that a baseline of length 16 wavelengths traverses an interval of $\Delta w=\pm4$ over the
course of a 2-hour observation, gridding in $w$ can substantially reduce integration time
within time bins.
% need a better solution to this problem!

The frequency dependence of an array's primary-beam response and $uv$-sampling 
allows visibilities at different frequencies to be coherently integrated for different
amounts of time.  Frequency-dependent gridding can substantially complicate delay-spectrum analysis, 
where one would prefer to
treat measurements uniformly at all frequencies within the band used in the delay transform, and ideally
over even wider bandwidths, in order to improve the performance of the wide-bandwidth, delay-domain
cleaning described in \S\ref{sec:implementation:cleaning}.
For this reason, we examine a two-stage gridding process, wherein entire frequency spectra for baselines are first
accumulated into time bins on the basis of the the narrowest coherence interval within the band (typically
at the highest frequencies).  The results of this first stage of accumulation can then be used for
wide-bandwidth cleaning and smooth foreground extraction.  At this point, only measurements that add in phase
will have been accumulated, but some frequencies, particularly the lower ones, will not have been accumulated
for as long as they might.  Since the 21cm reionization signal will generally be examined within $\sim\!$8-MHz
sub-bands over which the cosmological signal is not expected to evolve 
substantially, one can achieve better sensitivity 
by further accumulating each of these smaller cosmological sub-bands
into $(u,v,w,t)$-bins using coherence intervals
determined only for the frequencies within the sub-band.  We treat samples at
different frequencies within a sub-band uniformly, as the difference in coherence length
is relatively minor within a sub-band.

It is worth noting that the gridding scheme presented here for the delay-spectrum technique
differs substantially for other schemes that have been recently presented in the literature 
(e.g. \citealt{bowman_et_al2009,vedantham_et_al2012}) in
that within an 8-MHz sub-band, every frequency channel has been formed from the same weighted sum 
of constituent baselines.  That is, two baselines must sample the same $(u,v,w,t)$-bin at
the same frequency to be added together; we do not combine measurements of the same bin that
come from different frequencies.
In essence, binning along $(u,v)$
within each sub-band is done according to a physical baseline length (e.g. meters), rather than
the frequency-dependent units that $(u,v)$ represent.

\section{Simulating Foregrounds and Reionization in Delay Spectra}
\label{sec:simulation}

In order to examine in detail how foregrounds behave under the delay transformation,
we construct a sky model that consists of the foreground components discussed below, emulating both
their angular distribution on the sky and their expected evolution versus frequency.  We produce
simulated measurements of this sky by applying the response of our current best model of the 
PAPER primary beam and summing across the spatial response of the frequency-dependent fringe pattern for 
baselines of lengths 16, 32, 64, and 128 wavelengths at 150 MHz.  To the resultant visibility spectra,
we add Gaussian noise corresponding to a predicted frequency-dependent system temperature.
The system temperature used assumes contributions
from the galactic synchrotron sky temperature ($\sim\!400$ K at 150 MHz in a colder
region of the sky) and a flat-spectrum, 100-K receiver temperature.  This system temperature is
then integrated down using integration time and contributions from multiple baselines to a noise
level of a fiducial 132-antenna, 200-day PAPER observation.  The details of such
an observation are detailed in P12, although that predicted sensitivity has been modified here so baselines only contribute to measurements at
$k$ values that, on the basis of
the simulation work in this section, they do not exhibit 
foreground contamination.  The modified sensitivity curve is 
illustrated with a dotted line in Figure \ref{fig:sensitivity_with_fg}.

In this simulation, we assume ideal bandpass calibration.  In practice, 
the availability of strong calibrator sources and the engineered smoothness of passbands on the 
spectral scales of interest make departures from this assumption of negligible consequence.
Of greater concern is the possible presence of unmodeled departures from smooth spectral responses
introduced through
imperfections in the analog signal path such as cable reflections and cross-talk.  For example,
cable reflections can introduce an echo of the original signal at a time delay much larger than
is geometrically possible for a given baseline length, effectively scattering smooth-spectrum
foreground emission to much higher delay modes.
However, addressing
systematics such as these are beyond the scope of this paper.

Finally,
we flag data to imitate the flagging used to excise RFI in observations from PAPER's deployment
in South Africa.  
The distribution of flagging is modeled from actual observations.
The most substantial flagging occurs in a band near 137 MHz that is persistently occupied
by transmissions from the Orbcomm satellite constellation.

\subsection{Model Components}

\subsubsection{Galactic Synchrotron Emission}

Our simulation includes a modeled contribution from galactic synchrotron
emission, including both the expected spectral and spatial variation of the
signal.  The spatial structure of galactic synchrotron emission was taken
from the all-sky continuum survey at 408 MHz by \citet{haslam_et_al1982}.  This
map was extrapolated to frequencies in the 100-200 MHz band using a $\nu^{-2.5}$
power-law of brightness temperature versus frequency, making the explicit
assumption that synchrotron emission on the angular scales of interest varies uniformly
with frequency.  Using this sky model, we apply the simulated primary beam response of
PAPER antennas at 150 MHz and simulate the phase and amplitude of
visibilities measured by baselines of the lengths listed above.
These simulated
visibilities include the effect
of a baseline's changing $uv$-sampling versus frequency, as well as the changing sky
brightness versus frequency.  

To improve the computational tractability of the simulation,
the frequency dependence of the primary beam is not included for synchrotron emission, although it is
for the point-source foreground component described below.  While the spectral variation of the primary
beam impacts how foregrounds appear under the delay-spectrum transformation,
we argue that the point-source component of this simulation suffices to test the interaction of foregrounds
with the spectral and spatial variation of the primary beam.  The galactic synchrotron component of the foreground
simulation is instead designed to explore how the frequency dependence of $uv$-sampling interacts with a
steep-spectrum foreground with spatial structure that decreases rapidly toward smaller angular scales.

Our fiducial observations are for east-west oriented baselines measuring
a colder patch of sky centered near $\alpha=$03:00, $\delta=$-30:00, with observing parameters
chosen to match those of the PAPER deployment in South Africa at JD2455746.8.

\subsubsection{Point Sources}

A model point-source foreground is generated assuming random placement on the sphere.
Source counts were generated from a power-law in source counts versus source strength at 150 MHz, with 
a power-law index of -2.0, normalized
to one 100-Jy source per Jy flux interval per 10 sr.  This distribution was determined empirically from
pixel counts in PAPER maps, and yields a total of 10,000 sources with strengths from 0.1 Jy to 100 Jy.
Each source's frequency spectrum was also modeled as a power-law, with spectral index drawn from a normal
distribution centered at -1.0, with a standard deviation of 0.25.
Each source spectrum was multiplied by a model of the frequency-dependent primary beam response of 
PAPER antennas at the source's topocentric position. 

Ionospheric refraction was included for each source, with an apparent angular offset at 150 MHz drawn from
a zero-centered normal distribution with standard deviation of 1 arcminute.  We use a single
angular offset per source; since the
delay transform is linear, a single net angular offset models the offset averaged over many integrations.  Using
the full angular offset with RMS of 1 arcminute demonstrates a worst-case model of ionospheric 
refraction.  Apparent offsets were
extrapolated to other frequencies using
the characteristic $\nu^{-2}$ dependence of refraction angle versus frequency.  Frequency-dependent
refraction angles introduce a phase offset versus frequency whose magnitude is determined by the length
of the projection of a baseline perpendicular to the source direction.
Generally, the smooth variation of phase across the spectrum resulting from ionospheric refraction is expected
to have only a minor effect on the breadth of the delay spectrum of a source; we include it in the simulation for
completeness.

\section{Results}
\label{sec:results}

\begin{figure}\centering
    \includegraphics[width=.75\columnwidth]{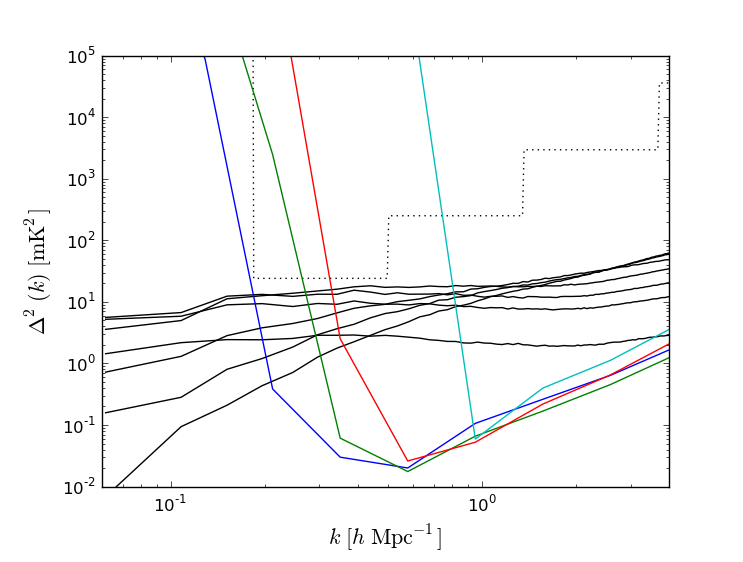}\caption{
The predicted PAPER-132 power-spectrum sensitivity from P12 (dotted black), re-plotted to show binning intervals
in $k$, and modified to only include contributions at each $k$ value
from baselines that do not show foreground corruption there, is overlaid on noiseless simulations
of contamination by smooth-spectrum foreground
emission observed by baselines of length 16 (blue), 32 (green), 64 (red), and 128 (cyan) wavelengths
at 150 MHz.  See
\S\ref{sec:simulation} for simulation details.
Simulated 21cm reionization power spectra from \citet{lidz_et_al2008}, with 
ionization fractions of
0.02, 0.15, 0.21, 0.54, 0.82, 0.96, (black curves, top to bottom, respectively, on plot right)
are also shown, projected to redshift 7.9, corresponding to 160 MHz.
It should be noted that these predicted power spectra are considerably more pessimistic than
those used in \citet{beardsley_et_al2012}.
Judicious use of windowing functions produces fall-off in foreground emission at a characteristic
scale that depends on baseline length.  The rise of foreground emission at higher $k$ is not noise,
but rather is
the result of sidelobes of smooth-spectrum foreground emission that are not fully suppressed by
the windowing and deconvolution steps used to compute the delay spectrum, multiplied by the
$k^3$ factor in $\Delta^2(k)$.
}\label{fig:sensitivity_with_fg}
\end{figure}

Applying the delay-transform foreground isolation technique described in \S\ref{sec:dly_transform} and
\S\ref{sec:implementation} to the simulated visibilities described in \S\ref{sec:simulation}, we produce
the results illustrated in Figure \ref{fig:sensitivity_with_fg}.  In this figure, we see that foregrounds are
effectively isolated below $k\sim0.2h\ {\rm Mpc}^{-1}$ for baselines of length $|\vec u|<16$.
This cutoff lies above the inherent cutoff imposed by the 8-MHz bandwidth used in the delay transform,
which falls at $k\approx0.05h\ {\rm Mpc}^{-1}$, and reflects an added breadth of foreground contamination
that results from the changing baseline response versus frequency that is explicitly addressed in the
delay-transform technique.

\begin{figure}\centering
    \includegraphics[width=.95\columnwidth]{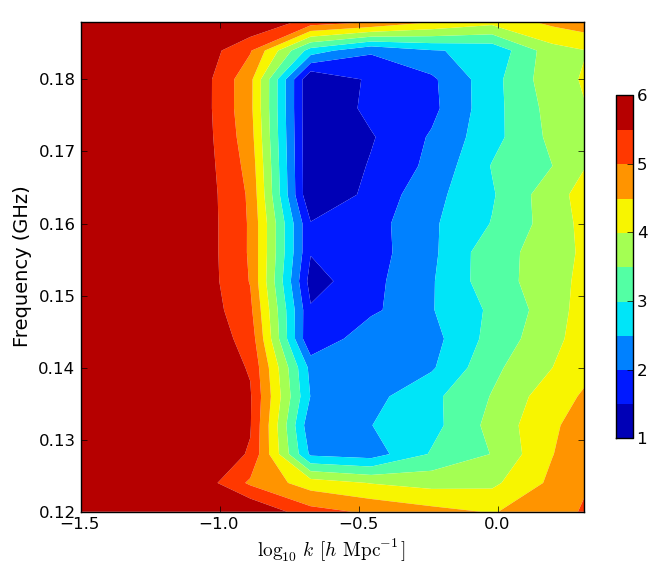}\caption{
$\Delta^2(k)$ versus frequency and $\log_{10}(k)$, computed using the delay transform,
 for simulated measurements of 
a 16-wavelength baseline at 150 MHz.  Different frequency values denote the centers of the various 8-MHz
sub-bands over which the delay
transform is applied.  As detailed in \S\ref{sec:simulation}, these simulations
include the effects of noise, data flagging, and foreground emission consisting of galactic
synchrotron emission and extragalactic point-sources with power-law spectra.
Color denotes $\log_{10}({\rm mK}^2)$, ranging from 1 (blue) to 6 (red).  
In this plot, the horizon limit on the geometric delay of celestial emission is clearly seen
in the rapid fall-off of bright foreground emission versus $k$.  The subsequent rise in temperature
versus $k$ is the result of $k^3$ dependence of $\Delta^2(k)$ multiplying flat thermal noise.
}\label{fig:k3pk_vs_k_vs_fq}
\end{figure}

\begin{figure}\centering
    \includegraphics[width=.95\columnwidth]{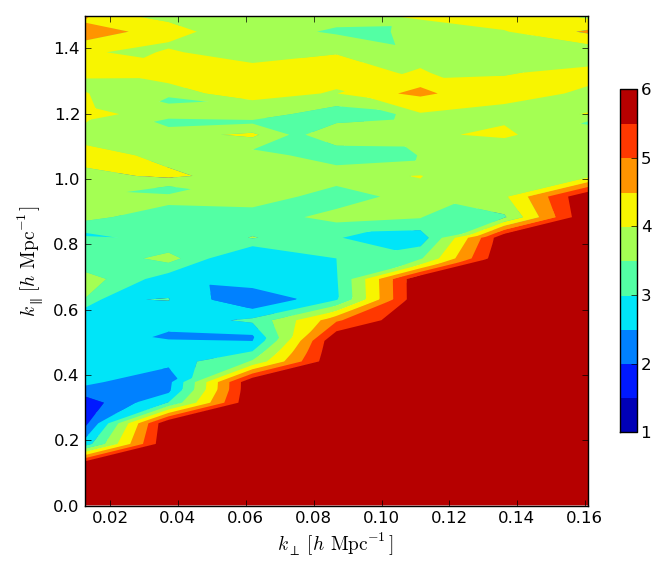}\caption{
$\Delta^2(k)$ versus $k_\perp$ and $k_\parallel$, derived from the same simulations
as in Figure \ref{fig:k3pk_vs_k_vs_fq}, but using a single delay-transform band centered at 150 MHz.
Color denotes $\log_{10}({\rm mK}^2)$, ranging from 1 (blue) to 6 (red).  
In this plot, the broadening of foreground contamination at lower $k_\parallel$ 
increases as a function of $k_\perp$, reflecting how the horizon limit
in delay-space increases as a function of baseline length.  The ``wedge'' of foreground contamination illustrated
here parallels the simulated foreground contamination found in 
\citet{datta_et_al2010} and \citet{morales_et_al2012}, and is indicative of the fundamental shape
of smooth-spectrum foreground emission as observed by an interferometer that has a
frequency-dependent sampling of the $uv$-plane.
    }\label{fig:k3pk_vs_kperp_vs_kpara}
\end{figure}

Given the
$k$-domain extent of smooth-spectrum foregrounds and the inherent frequency dependence of the PAPER primary beam,
we find that the mode-mixing resulting from the projection of delay-modes along the angular direction
substantially compromises measurements at smaller $k$-modes, particularly for longer baselines.  However,
the effect of this mode-mixing is still bounded by the maximum geometric delay for
a baseline of a given length.  The inherent breadth of typical power-law source spectra, the angular distribution
of point-source and galactic synchrotron foregrounds, refraction
through the ionosphere, and the simulated response of the PAPER primary beam do not appear to substantially broaden foreground contamination
beyond this limit.
The dependence of this limit on frequency is illustrated in Figure \ref{fig:k3pk_vs_k_vs_fq}, and its dependence
on baseline length is illustrated in Figure \ref{fig:k3pk_vs_kperp_vs_kpara}.

Following the calculations in P12 for the sensitivity of a 132-element array in a 
configuration consisting of 12 columns of 11
closely-packed antennas, accounting for the fact that baselines
cannot contribute sensitivity at $k$ values where they show foreground contamination, and binning in
intervals of $\log k=1$, we find that
132 PAPER antennas achieve adequate sensitivity in 200 days of
observation to constrain fluctuations of 30 mK$^2$ at the $3\sigma$ level near $k\approx0.2h\ {\rm Mpc}^{-1}$.
As shown in Figure \ref{fig:sensitivity_with_fg}, the foreground isolation obtained through the delay
transform combines with the predicted sensitivity curve (scaling as $k^{5/2}$) to
open a window on shorter baselines for accessing $k$-modes of reionization
that are uncorrupted by smooth-spectrum foregrounds near $k\approx0.2h\,{\rm Mpc}^{-1}$. 

\begin{figure}\centering
    \includegraphics[width=.95\columnwidth]{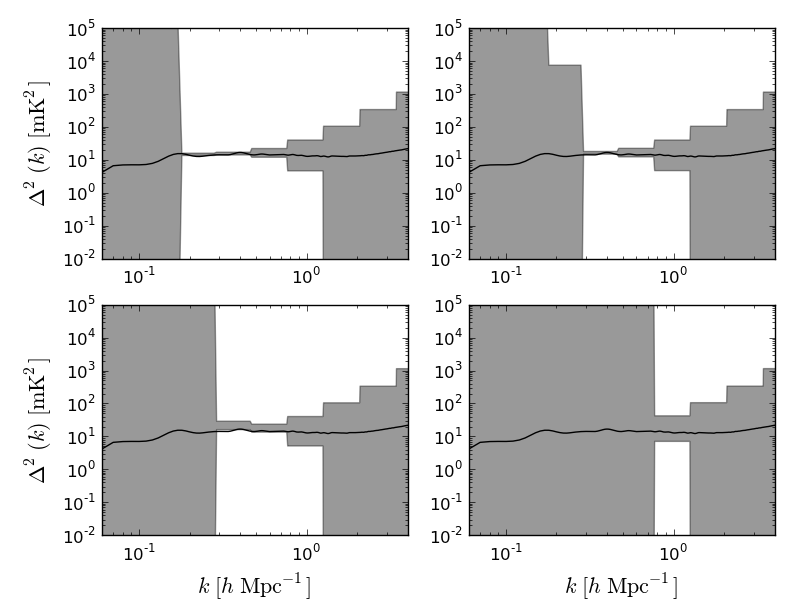}\caption{
The measured $\Delta^2(k)$ versus $k$ with shaded 2$\sigma$ error bars for baselines
of length 32 (upper-left), 64 (upper-right), 128 (lower-left), and 256 (lower-right) meters, 
corresponding to 16, 32, 64, and 128 wavelengths, respectively,
as measured using the delay spectrum technique.
Error bars include contributions from foregrounds and from thermal noise, assuming observations with a future
array 100 times more sensitive than a 132-antenna PAPER array observing for 200 days.  The dark solid line
denotes the true $\Delta_{21}^2(k)$ power spectrum used in the simulation.
    }\label{fig:final_eor_errorbar}
\end{figure}

Extending these simulations to a future array with identical antenna properties but with 100 times the 
sensitivity, we see in Figure \ref{fig:final_eor_errorbar} that
the measured power spectrum of 
21cm EoR fluctuations reproduces the fiducial input model despite the broadening of the PSF shown
in Figure \ref{fig:pk_resp_of_dly_bin}
that results from the frequency dependence of baseline length.  The original input model falls within
the 2$\sigma$ error bars at all measurements, illustrating that the delay-spectrum approach can be effective
not only in a first-detection scenario, but also for characterizing 21cm reionization with a next-generation
instrument.

\section{Discussion}
\label{sec:discussion}

The technique we have presented bears broad
similarities to other approaches described in the literature 
\citep{morales_hewitt2004,bowman_et_al2009,liu_tegmark2011} in that it
seeks to exploit spectral smoothness in foreground removal, and in that it contends with
the inherent frequency dependence of $uv$-sampling.  The technique we have described
differs, however, in its light dependence on synthesis imaging.  Instead,
the focus is on foreground removal per-baseline on the basis of spectral smoothness.  While image-domain foreground
modeling will undoubtedly still play an important role in calibration and bright source removal,
we have shown that by using delay-domain foreground separation, the stringent calibration requirements required for
adequate foreground removal via synthesis imaging are 
relaxed into constraints on spectral and spatial smoothness in instrumental response.

Our approach appears to be quite similar to the independently derived approach recently described 
in \citet{vedantham_et_al2012}, which
involves the application of the Chirp Z Transform to interferometric measurements that are gridded in units of
physical length, rather than wavenumber.
This approach effectively yields the delay transform we describe, and we reach similar 
conclusions regarding the level of foreground removal that is possible with the effective use of windowing
functions in the line-of-sight Fourier transformation.
An important distinction between our approach and theirs
is our emphasis on the per-baseline application of this technique.  Using the insight that
the transforms described in each paper can be physically interpreted as yielding the geometric delay in an 
interferometer pair, we find that imaging is altogether unnecessary for the effective application of this technique.
In fact, one of the significant advantages of our technique is that data may be retained in the native
gridding produced in the interferometer, thereby avoiding ``gridding contamination,'' 
inter-channel ``jitter,'' confusion effects, and
many of the other complications described in \citet{vedantham_et_al2012}.  Futhermore, the independent treatment
of each baseline in our approach allows the dependence of foreground contamination on baseline length
to be addressed by only using shorter baselines to constrain the wavemodes where longer baselines remain contaminated.

Understanding the relationship between delay-space and $k$-space also provides a
simple interpretation of the foreground ``wedge" described in \citet{datta_et_al2010} and
\citet{morales_et_al2012},
and is seen in Figure \ref{fig:k3pk_vs_kperp_vs_kpara}.  Foreground
emission is bounded by a maximum delay in delay space, which corresponds to a maximum $k_\parallel$
in $k$-space.  The value of this maximum delay depends on baseline length,
with foreground emission extending to higher $k_\parallel$ modes on
longer baselines.  Longer baseline lengths correspond to higher $|k_\perp|$ modes, creating the 
linear trend in $k_\parallel$ versus $k_\perp$ of foreground contamination that is the wedge
shape.  Sources near the field edge enter at higher delays, thus mapping them
to higher $k_\parallel$ modes than sources near field center, as stated by \citet{morales_et_al2012}.
The wedge can then be understood, not as resulting from
foreground model errors themselves, but as an intrinsic shape of foreground
emission, the residuals of which will remain if there is imperfect foreground
subtraction.

An important consequence of the viability of foreground removal on a per-baseline basis is that
it allows somewhat more flexibility in array configuration.
Notably, this foreground removal technique is viable for the maximum-redundancy array
configurations advocated in P12 that can dramatically enhance sensitivity.  As shown in
Figure \ref{fig:sensitivity_with_fg}, the combination of these techniques opens
a substantial window for accessing $k$-modes of reionization
that are uncorrupted by smooth-spectrum foregrounds.  

A notable foreground component that has not been included in the discussion in this paper is polarized galactic
synchrotron emission.  As has been noted many times in the literature, linear polarization measurements of 
emission with high rotation measure pick up the rapid rotation of polarization angle with frequency that
is of concern for measuring 21cm reionization \citep{furlanetto_et_al2006,jelic_et_al2008,pen_et_al2009,geil_et_al2011}.  This is
illustrated in the equation for Faraday rotation of polarization angle,
\begin{equation}
Q+iU = (Q_{\rm in}+iU_{\rm in})e^{-i {\rm RM}\,\lambda^2},
\end{equation}
where ($Q_{\rm in}$, $U_{\rm in}$) are the intrinsic polarization parameters of a source, and ($Q$,$U$)
are their observed values.
Faraday rotation effectively modulates otherwise smooth-spectrum synchrotron
emission by a ${\rm RM}\,\lambda^2/2\pi$ delay term that can scatter smooth-spectrum emission beyond the 
maximum geometric delay at the horizon that was derived in \S\ref{sec:dly_transform}.

We are currently exploring how the delay-spectrum technique can be extended to handle polarized foregrounds.
Managing polarized foregrounds will begin with the careful formation of the Stokes $I$ polarization parameter,
but imperfect polarization calibration will leave a residual, direction-dependent leakage of linear polarization
into $I$.  In a future publication, we will explore how the whole-bandwidth delay-domain deconvolution 
from \S\ref{sec:implementation}
might be extended to employ rotation measure synthesis \citep{brentjens_debruyn2005} to model and remove
polarized foregrounds that add coherently over much wider bandwidth than the peak 21cm reionization signal is
expected to occupy.

\section{Conclusion}
\label{sec:conclusion}

The delay-transform methodology we have described applies separately
to each baseline, concentrating smooth-spectrum foregrounds within the bounds
of the maximum geometric delays for each baseline.  
The measured signal outside of these modes is dominated by sidelobes of unsmooth
emission on the sky, such as the predicted 21cm signature of reionization. The
response of delay-modes is peaked toward specific line-of-sight
components of the 3D wavevector $k$, making them effective
probes of the 21cm reionization power spectrum.
The primary complication
inherent to this approach, relative to extracting strictly line-of-sight
fluctuations with overlapping $uv$-coverage generated by larger future arrays, is
that extracted $k$-modes have non-negligible projections along image-domain
angular coordinates. As a result, confining smooth-spectrum foregrounds to low
k-modes requires that instrumental response be smooth both spectrally and
spatially.

PAPER is unique among current experiments targeting cosmic reionization in its pursuit
of antenna elements with precisely these characteristics. As a result,
PAPER may be well-positioned to explore and capitalize on this new approach for
measuring the three-dimensional power spectrum of reionization, and to quantify
the need for antenna elements with these characteristics.  Work in this
area will likely have repercussions
for the design of future reionization experiments such as the Hydrogen Epoch of Reionization
Array (HERA) proposed in a white-paper to the Astro2010 Decadal Survey.

PAPER is currently pursing this
strategy as its primary path to detecting reionization with 132 deployed antennas.
PAPER should achieve adequate sensitivity in 200 days of
observation to constrain fluctuations of 30 mK$^2$ at the $3\sigma$ level near $k\approx0.2h\ {\rm Mpc}^{-1}$.
Not surprisingly, the crux of successfully applying delay-spectrum foreground isolation at these
sensitivity levels
will be mastering frequency-dependent instrumental systematics such as reflections,
resonances, or cross-talk in the analog signal path.

% --------------------------------------------------------------------------
% --------------------------------------------------------------------------
% --------------------------------------------------------------------------

\bibliographystyle{apj}
\bibliography{biblio}

% Figures

%\begin{figure}\centering
%    \includegraphics[width=.75\columnwidth]{plots/kpara_vs_umag.png}\caption{
%The above plot illustrates the slope in the $k_\parallel$ response of a delay-transform bin across
%the FoV that results from the inherent scaling of $uv$-coverage with frequency.  Plotted are baselines
%with lengths in wavelengths at 150 MHz of 16 (blue), 32 (green),  64 (red), and 128 (cyan) for delay bins
%centered at $k_\parallel=0.25$, $0.5$, and $1.0\ h\ {\rm Mpc}^{-1}$ (dotted, dashed, and sold plots, respectively).  
%The inner
%vertical black lines illustrate the approximate radius in $|(l,m)|$ of the FoV of the primary beam of
%a PAPER antenna, illustrating the approximate smearing in $k_\parallel$ response that results from the sloping
%response pattern.  The outer vertical black lines similarly illustrate the absolute limits on smearing
%imposed by limiting the FoV at the horizon.  For $|\vec u|\sim100$ at 150 MHz, smearing mixes modes
%over an interval of $\Delta k_\parallel\approx 0.01\ h\ {\rm Mpc}^{-1}$, and becomes a negligible effect
%at $k_\parallel>0.1$.
%    }\label{fig:kpara_vs_umag}
%\end{figure}

\end{document}